\documentclass{comnet}

\usepackage{graphicx,url}
\usepackage{multirow}
\usepackage[english]{babel}   
\usepackage[utf8]{inputenc}  
\usepackage{float}
\usepackage{comment}
\usepackage{color}
\usepackage{booktabs}
\usepackage{rotating}
\usepackage[dvipsnames]{xcolor}
\usepackage{subfigure}
\usepackage{amsmath,amsthm}

\begin{document} 

\title{Analysis of account behaviors in Ethereum during an economic impact event %Analysis of Ethereum account behavior during an economic impact event
%\footnote{Essa pesquisa é financiada por CNPq, CNPq/Amazon AWS (Processo 440069/2020-3), FAPEMIG, CAPES e UFJF.}
}

\author{
\name{Pedro Henrique F. S. Oliveira, Daniel Muller Rezende, Heder Soares Bernardino, Saulo Moraes Villela$^*$, Alex Borges Vieira}
\address{Computer Science Department, Federal University of Juiz de Fora (UFJF), Juiz de Fora, Minas Gerais, Brazil\email{$^*$Corresponding author: \{phfsoliveira,daniel.muller,heder\}@ice.ufjf.br, \{saulo.moraes,alex.borges\}@ufjf.edu.br}}
\and
\name{Glauber Dias Gonçalves}
\address{Computer Science Department, Federal University of Piau\'i (UFPI), Picos, Piau\'i, Brazil\email{ggoncalves@ufpi.edu.br}}
}

%\email{\{phfsoliveira,daniel.muller,heder\}@ice.ufjf.br, \{saulo.moraes,alex.borges\}@ufjf.edu.br}

%\email{ggoncalves@ufpi.edu.br}

\maketitle

\begin{abstract}
    {One of the main events that involve the world economy in 2022 is the conflict between Russia and Ukraine. This event offers a rare opportunity to analyze how events of this magnitude can reflect the use of cryptocurrencies. This work aims to investigate the behavior of accounts and their transactions on the Ethereum cryptocurrency during this event. To this end, we collected all transactions that occurred two weeks before and two weeks after the beginning of the conflict, organized into two groups: the collection of the accounts involved in these transactions and the subset of these ones that interacted with a service in Ethereum, called Flashbots Auction. We modeled temporal graphs where each node represents an account, and each edge represents a transaction between two accounts. Then, we analyzed the behavior of these accounts with graph metrics for both groups during each observed week. The results showed changes in the behavior and activity of users and their accounts, as well as variations in the daily volume of transactions.}
    {Ethereum, Flashbots, World Economy, Temporal Graphs}
\end{abstract}

\section{Introduction}~\label{introducao}
In April 2021, Russian armed forces carried out a military mobilization on the border of Ukraine, compared to the mobilization for the annexation of the Republic of Crimea to the Russian Federation, which took place in 2014\footnote{\url{https://www.reuters.com/article/us-ukraine-crisis-usa-idUSKBN2BV2Z3}}. Almost a year later, on February 24, 2022, Russia carried out a new military mobilization on the territory of Ukraine, starting a conflict between these two countries\footnote{\url{https://www.cnbc.com/2022/02/24/russian-forces-invade-ukraine.html}}. To avoid a further escalation of global tensions, the way that other countries around the world have sought to interfere is through sanctions on Russia and economic support for Ukraine\footnote{\url{https://www.bbc.com/news/world-europe-60125659}}. Events of great magnitude like this tend to involve financial transactions, which may reflect the use and popularization of cryptocurrencies, based on blockchain technology, for transfers and acquisition of digital assets in a decentralized way.

Among the various existing cryptocurrencies, the \textit{Ethereum} stands out as one of the leading platforms for trading cryptocurrencies, currently having the second largest market cap with over 363 billion dollars\footnote{\url{https://coinmarketcap.com/}}. Additionally, Ethereum is the main cryptocurrency that includes the functionality of storing and processing program codes – called smart contracts – within its blockchain. 
%Outra particularidade do Ethereum é a organização Flashbots, que desenvolveu um ecossistema digital que possibilita a um usuário da rede enviar conjuntos de transações (conhecidos como bundles) diretamente ao minerador.~\cite{Wehmuth2017} Da maneira usual na rede, uma transação seria lançada na mempool, para que ela pudesse ser selecionada para mineração, e então aprovada ou não. No caso de rejeição, a taxa de transação ainda assim deveria ser paga ao minerador. Devido ao procedimento diferente, questões como essa podem ser resolvida com transações do tipo Flashbots, que passam a ser uma alternativa válida para usuários que pretendem enviar transações de maneira mais privativa, rápida e com menores taxas. 

Faced with this situation, there are still many uncertainties about how such an armed conflict with impacts on the world economy and financial market can reflect on crypto-assets. 
An important question arises in this context: How to monitor large crypto platforms like Ethereum to observe the possible impact of events external to these platforms on the behavior of their users in the shortest possible period of time?

To begin to elucidate this issue, in this work we propose a characterization of the behavior of Ethereum accounts within a short period of time, specifically, less than a month.
In this regard, we collect all transactions on Ethereum between two weeks before and two weeks after the conflict started. Next, we contrast two different views about the behavior of the accounts involved in these transactions in this period. 
The first view represents a general context that includes all accounts that performed transactions in the four weeks collected. The second view represents a more specific context, contemplating only accounts with transactions on Ethereum via the service \textit{Flashbots Auction}, which is a new communication mechanism between platform maintainers (miners) and users\footnote{\url{https://docs.flashbots.net/}}. % idealizado pelo grupo de pesquisa \textit{Flashbots}\footnote{\url{https://docs.flashbots.net/}}.

%Para isso, modelamos uma rede principal por meio de um grafo direcionado que representa nele todas as transações realizadas no período, compondo as arestas do grafo, bem como as contas que as realizaram, compondo os vértices do grafo. Em paralelo, modelamos uma outra rede que é um subconjunto da rede geral. Esta é composta apenas por transações realizadas por meio de flashbots.

%O Flashbots Auction é o mecanismo de comunicação entre mineradores e usuários idealizado pelo grupo de pesquisa Flashbots\footnote{\url{https://docs.flashbots.net/}}, utilizado para a extração das receitas MEV de forma eficaz e segura.

The choice of the Flashbots service is due to its relevance as a mediating mechanism for transactions aimed at extracting MEV revenues, that is, the maximum profits obtained by trading crypto-assets. A relevant example is a scenario of buying and selling non-fungible tokens (NFTs). In it, users use techniques to monitor the network and profit from the price differences of these assets between exchange institutions or \textit{exchanges}. The analysis of Flashbots allows us to obtain an approximation of the behavioral characteristics of this aspect of the platform that has grown considerably in recent years.

In our characterization of Ethereum accounts with general and specific views, we explored properties of graphs, where vertices represent accounts and edges characterize transactions between two accounts. This approach allowed us to analyze the temporal dynamics of the degree of accounts, considering the value and number of transactions, as well as the identification of central accounts with anomalous behavior in a period that involves an event of impact on the economy.
To sum things up, this work has two main contributions: (i) a strategy to characterize the behavior of users on cryptocurrency platforms such as Ethereum, which allows the use of various services to trade cryptocurrencies within the platform, and (ii) the characterization of the behavior of the accounts due to an event external to the platform of relevance for analysis. We used the Flashbots service and the current conflict between Russia and Ukraine as a case study.
%no período próximo à invasão russa, demonstrando como características da rede geral e de transações flashbots se modificaram com o passar das semanas. 

The next sections of this article are organized as follows. 
We present related works in the Section \ref{trabalhos_relacionados}. In Section~\ref{metodologia}, we describe the collection of data used, while in Section~\ref{sec:modelagem} we describe the methodology for the analysis of these data. The results obtained are explored in Section~\ref{sec:framework} and, the conclusions and discussion of future work are presented in Section~\ref{conclusao}.

\section{Related works}~\label{trabalhos_relacionados}
Graphs have become an important way of modeling and characterizing transactions on cryptocurrency platforms. The first graph-based analyzes were conducted on Bitcoin, which was the first crypto-asset platform conceived, based on blockchain technology, and is still considered the largest crypto-asset currently in terms of the number of transactions and value of invested capital. In~\cite{miller2015discovering,sun2019bitvis} graphs were used to model the structure of the Bitcoin network to identify profiles of influential or suspicious accounts. However, Bitcoin is considered a first-generation crypto-asset that does not support smart contracts and the various contract-based services for trading crypto-assets.

Ethereum, on the other hand, is a second-generation cryptocurrency platform, in which the use of smart contracts was first conceived. Given the variability of services for contract-based cryptocurrencies, graph models have been widely developed for Ethereum. The model proposed in~\cite{Chen2020} makes use of graphs to represent creating smart contracts, invoking smart contracts, and performing external transactions. Through these graphs, the authors are able to make use of centrality metrics in complex networks to detect important and anomalous vertices in Ethereum.

Techniques that exploit temporal graphs have also been proposed in the literature for the analysis of the Ethereum platform. In~\cite{agarwal_detecting_2021}, for example, features extracted from targeted temporal graphs were used to develop a predictive model based on machine learning trained to identify profiles of malicious accounts within this platform.
Furthermore, in~\cite{zhao2021temporal,bai2020poster,zanelatto2020transaction} graph-based models were used to analyze the evolutionary behavior of Ethereum over time since its origin in 2014.

Unlike the above-mentioned graph-based works, in this work, we propose temporal graph models to analyze the Ethereum platform in a short period before and after a reference event. To our knowledge, our work is the first to explore graph metrics with this approach. Additionally, we contrast metrics extracted from a complete temporal graph and its subgraph that respectively represent the general Ethereum accounts in the period of interest and the accounts that interacted with the Flashbots service.

It is also important to mention that our work uses Flashbots as an approximation for the behavior of users on Ethereum that aim to exploit maximum profits from the speculation of digital assets. In~\cite{daian2020flash,capponi2022evolution} the risks that the use of crypto-asset arbitrage services such as Flashbots pose to the security of Ethereum are analyzed, which may negatively affect the functioning of the platform and the experience of users. However, these works do not explore graph metrics and temporal analysis with reference to an event of an impact on the economy.
\section{Data Description}~\label{metodologia}
%
%\subsection{Período de coleta}\label{subsec:interval}
\subsection{Data characterization}\label{subsec:datacharacter}
Russian troops invaded eastern Ukraine on 24/02. From this, we chose to subdivide our dataset into four weeks. Weeks 1 and 2 are defined by the initial and final blocks, respectively with the identifiers 14174989 and 14265470, characterizing the pre-invasion period. Moreover, weeks 3 and 4 composed of the initial and final blocks, respectively with the identifiers 14265812 and 14355747, define the post-invasion period.

For the analyses, we chose to characterize the behavior of the network from two points of view. The first refers to the general context, in which all transactions that occurred on the network during the collection period are used as a database. The second addresses a specific context in which we focus on transactions that interact with the Flashbots Auction service.

Transactions that interact with this mechanism, after undergoing an error check in a \textit{gateway} denominated MEV-Relay, are sent directly to miners, making them invisible on the network until mined. This, in addition to preventing the occurrence of attacks that try to steal these revenues, contributes to a better user experience in relation to the expense of fees for failed transactions, making Flashbots a relevant transaction intermediary mechanism for our analysis.

%Através disso, com base na teoria dos conjuntos, definimos como o conjunto A, o grupo que engloba todas as transações ocorridas dentro do intervalo de coleta e como conjunto $C_2$ a parcela dessas transações que é do tipo Flashbots, de forma que B $\in$ A e A $\in$ U, com U sendo o conjunto universo que contém todas as transações históricas do Ethereum.

Amidst the collected data, we define as transaction volume all the trades successfully processed on the Ethereum network during the collection period. Therefore, we define accounts as all individual addresses that participate in a transaction, whether they are senders or recipients. The set of accounts that make up our database encompasses both smart contracts and user wallet addresses, not being necessary to differentiate them for the construction of account volume analyses. The distinction between contract and user was made by identifying addresses that do not have a public tag on Etherscan in the section~\ref{subsec:topaccs}.

\subsection{Form of collection}\label{subsec:collection}
To form our database, we collected all transactions carried out on the Ethereum blockchain between 02/10/2022 and 03/10/2022, which belong to the closed interval defined between blocks 14174989 and 14355747. This time interval was chosen in order to obtain behavioral information from the network before and after the Russian invasion of Ukraine. 
Transactions were collected through APIs Etherscan.io\footnote{\url{https://etherscan.io/}} and blocks.flashbots.net\footnote{\url{https://blocks.flashbots.net/}}. The first was used to obtain all transactions that took place on the Ethereum network during the collection period, being the source for obtaining the main characteristics of each transaction. The second was used to identify Flashbot-type transactions.

Both APIs have different purposes, while Etherscan.io provides all public data of Ethereum transactions, blocks.flashbots.net provides information aimed at extracting MEV revenues, such as the amount of ether transferred to coinbase, transaction fees, and total miner profits. For the purpose of this work, we focused our efforts on obtaining general data that allows the characterization of the network without delving into the field of arbitration and extraction of MEV revenues. Both datasets: general and Flashbots-type transactions, have the same collected attributes. Table \ref{tab:table1} shows the data used for the analyses.

\begin{table}[!ht]
\centering
\caption{Characteristics extracted from Ethereum transactions.}
\label{tab:table1}
\begin{tabular}{ll}
\toprule
\multicolumn{1}{c}{Attribute} & \multicolumn{1}{c}{Description} \\
\midrule
hash                                  & Hash code that identifies each transaction \\
blockNumber                           & Block number (required for sorting) \\
timestamp                             & \textit{Timestamp} setting the date the transaction was processed \\
to                                    & Transaction recipient account \\
from                                  & Transaction sender account \\
\bottomrule
\end{tabular}
\end{table}

\section{Modeling}\label{sec:modelagem}
To perform a network characterization, we initially model our problem using directed temporal graphs to represent two sets within the main Ethereum network: the set $C_1$, with all external transactions collected in the period from 02/10 to 03/10 ; and the $C_2$ set, with all transactions that interact with Flashbots Auction within the same time frame.

\begin{figure}[!ht]
        \centering
      	\includegraphics[page=1,width=.6\textwidth]{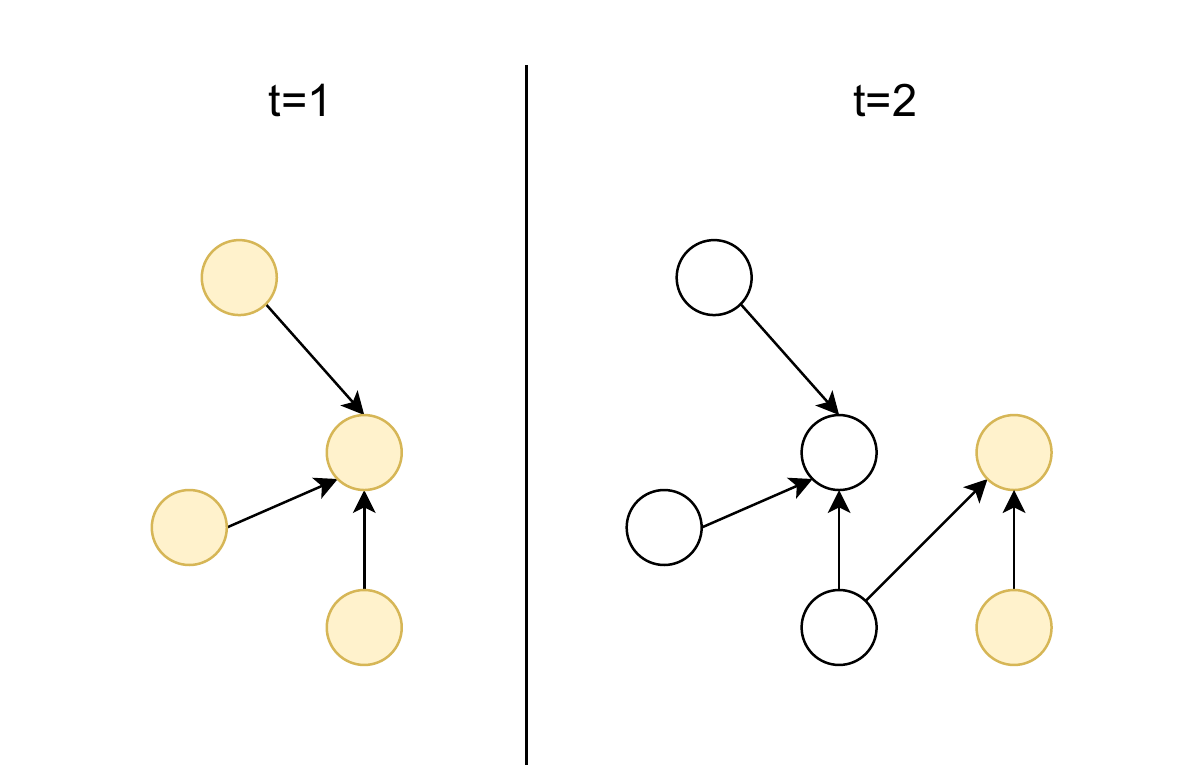}
        \caption{Example of a directed temporal graph.}
        \label{fig:tempgraph}
\end{figure}

Let a static graph $G$ be defined as a tuple $(V,A)$, with $V$ as a set of vertices and $A$ a set of edges, where $A \subseteq \ $\{$(u,v) \ | \ u, v \in V$\}. Since $(u,v)$ is an ordered pair, G is also considered a directed graph. Therefore, a directed temporal graph is defined as a sequence of directed static graphs, as a function of a discrete variable $t$. Figure~\ref{fig:tempgraph} shows a diagram representing a directed temporal graph. The circles represent the vertices of the graph, being filled in as they are added to the previous graph.

In the case of this work, temporal graphs will be used in two different instances: Analysis of main accounts and Analysis of new accounts.

\subsection{Analysis of main accounts}\label{subsec:mainacc}
In this situation, the discrete variable $t$ represents different weeks on which we focus our analyses. Therefore, $G_1$ represents a graph in the first week within the analysis interval; that is the week that includes the first block of the 10/02 to the last block of the 16/02. Likewise, $G_2$ forms a range from the first block on 02/17 to the last block on 02/23, and so on.

To assess the relevance of a given account in a period we focus on the degree metric. A degree represents in this problem the number of transactions that an account receives or sends, that is, the number of edges that connect with the vertex in question. We also separate the indegree metric, which considers only transactions with the account in question being the destination of the transaction, and outdegree, considering only transactions with the account in question being the origin of the transaction.

We understand that high degree accounts or accounts that show high degree growth from one week to the next show up as relevant accounts in the short/medium term, and therefore, will be the object of greatest interest in this work.

\subsection{Analysis of new accounts}\label{subsec:newacc}
For this scenario, the variable $t$ represents days that have passed since the first day. Thus, $G_1$ represents a graph containing all blocks and transactions that entered the blockchain on 02/10, $G_2$ all blocks and transactions that entered the blockchain on 02/11, and so on.

The objective of this approach is to analyze the volume of accounts involved in transactions each day and how many of these accounts had already been observed with recent activity on the network. To define a starting point, we collected all accounts that had appeared in transactions on the network between 01/10 and 02/10 and inserted them as a starting point in $G_1$, adding accounts that had already participated to the temporal graph. of recent transactions.

Let $N_t$ be the number of vertices that the temporal graph has on the day $t$. For each day $t$, we have that the daily volume of new accounts is defined by $N_{t}$ – $N_{t-1}$. For day 1, the number of new accounts is arbitrarily set to $N_1$. In this context, the volume of active accounts in a day $t$ is defined by the number of vertices of $G_{t}$ that have at least one transaction with \textit{timestamp} which corresponds to the day $t$.

\section{Characterization}~\label{sec:framework}
In this section, we present the results found for the analyzes carried out on the Ethereum network of transactions in the period described in Section \ref{metodologia}. We used the modeling techniques presented above to analyze the volume of transactions, and the activity of the accounts and later elaborate on the ranking of the nodes based on the change of degree between the weeks. As this is a characterization work, we focused on presenting the results without worrying about understanding the causes and motivations for them.

\subsection{Transaction Volume}~\label{subsec:txs}
\begin{figure}[!ht]
    \subfigure[Ethereum network transactions]{\includegraphics[page=1,width=.5\textwidth]{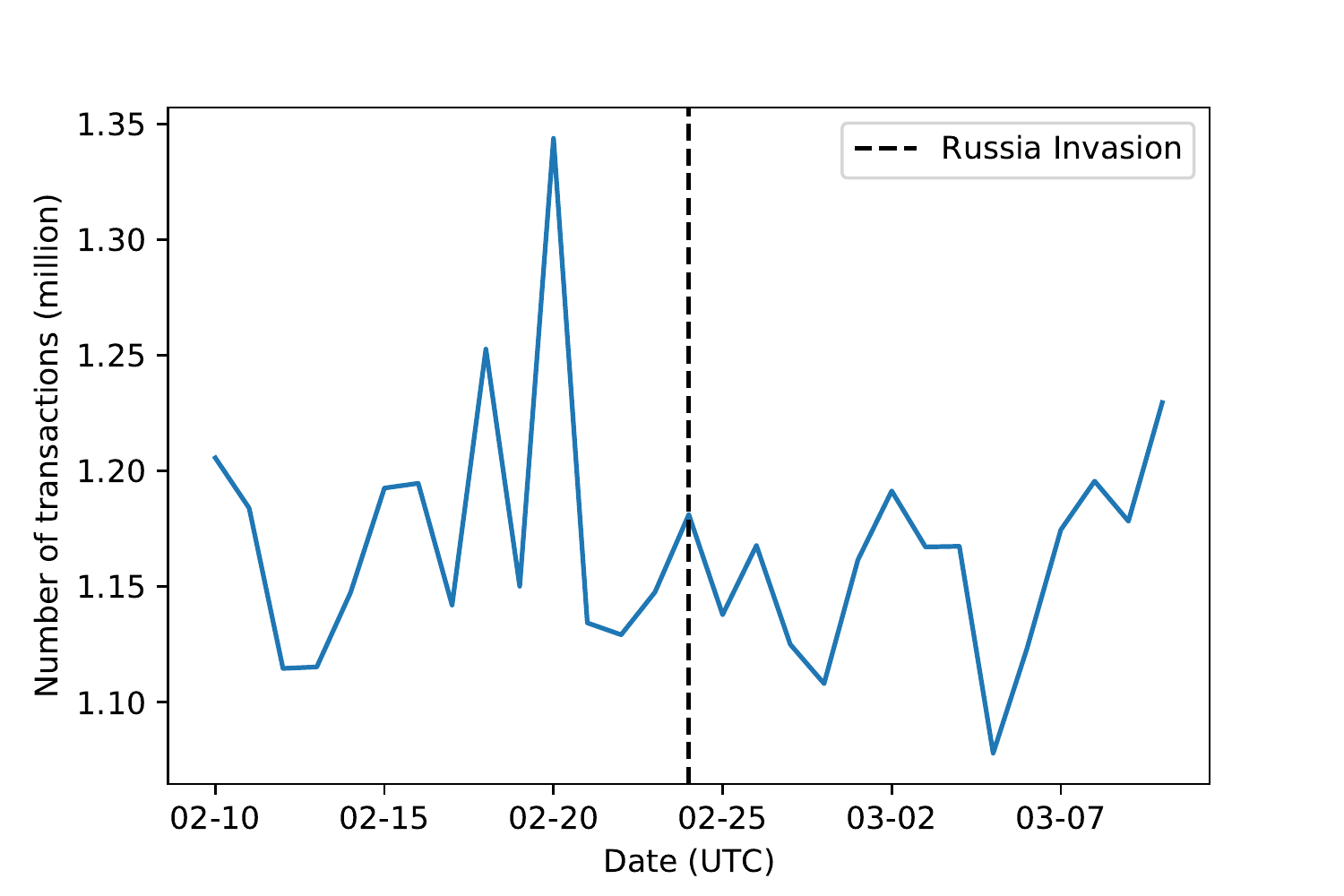}\label{subfig:txmain}}
    \subfigure[Flashbot Transactions]{\includegraphics[page=1,width=.5\textwidth]{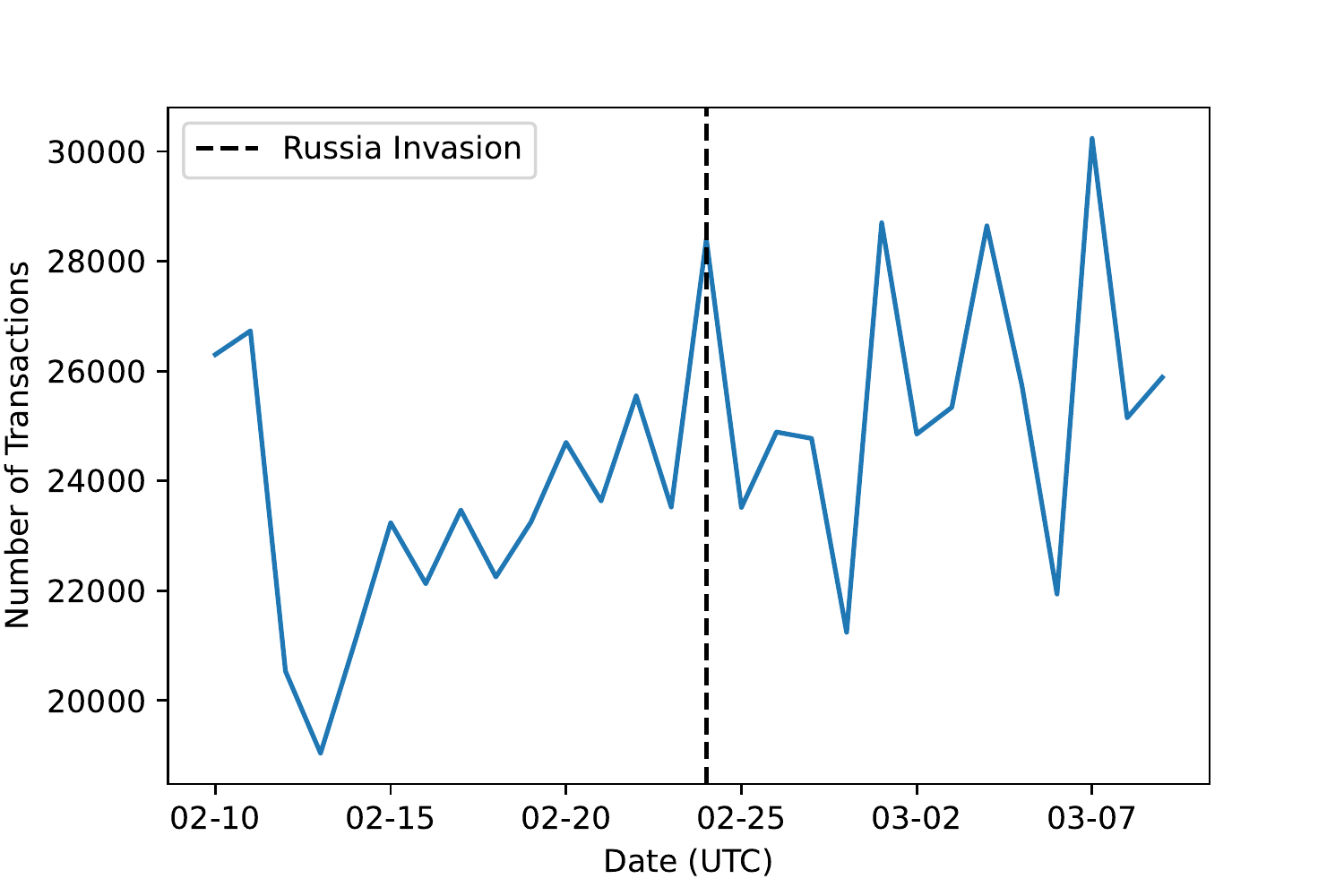}\label{subfig:txflashbots}}
    %\begin{subfigure}{0.5\textwidth}
     %   \centering
      %  \includegraphics[page=1,width=\textwidth]{grafs/flashbots_number_of_transactions_now.pdf}
      %  \caption{Transações diárias de Flashbots}
      %  \label{fig:tempgraph}
    %\end{subfigure}
    %\begin{subfigure}{0.5\textwidth}
    %    \centering
    %  	\includegraphics[page=1,width=\textwidth]{grafs/general_tx_num.pdf}
    %    \caption{Transações diárias da rede Ethereum}
    %    \label{fig:tempgraph2}
    %\end{subfigure}
    \caption{Daily volume of transactions.}
    \label{fig:txfig}
\end{figure}

In Figure \ref{fig:txfig}, we present the results obtained for the volume of network transactions. For this, we build time charts relating each transaction to the date on which it occurred to obtain the daily total of trades. As previously described, we have separated the analyzes into two points of view, the first as an overview of the network and the second in the context of Flashbots. With this in mind, it is worth noting that the graphs in Figure \ref{fig:txfig} are on different scales, so that the curves of each must be analyzed separately.

First, we analyze the set $C_1$, referring to the general behavior of the system, represented by Figure \ref{subfig:txmain}. In it, we can observe a clear difference between the periods before and after the invasion, in which in mid-February, the network had large peaks in transactions, the most significant being approximately 1.35 million close to February 20th. On the other hand, it is possible to highlight an evident drop in this number after the 20th, causing the total transactions in one day to reach a value of less than 1.10 million at the beginning of March.

Subsequently, we performed the same analysis for the set $C_2$, referring to Flashbots transactions, which are represented by Figure \ref{subfig:txflashbots}. Unlike the general context, the set $C_2$ shows a growth behavior throughout the analyzed period, having a more intense fall only between the 10th and 15th of February, causing the total number of negotiations to reach a value of less than 20 thousand. It is also worth noting that after the invasion, although the volume of transactions continued to grow, the fluctuations were more significant than before February 24, with variations of more than 6 thousand transactions.

\subsection{Account Activity}~\label{subsec:accounts}
\begin{figure}[!ht]
    \subfigure[Ethereum network accounts]{\includegraphics[page=1,width=.48\textwidth]{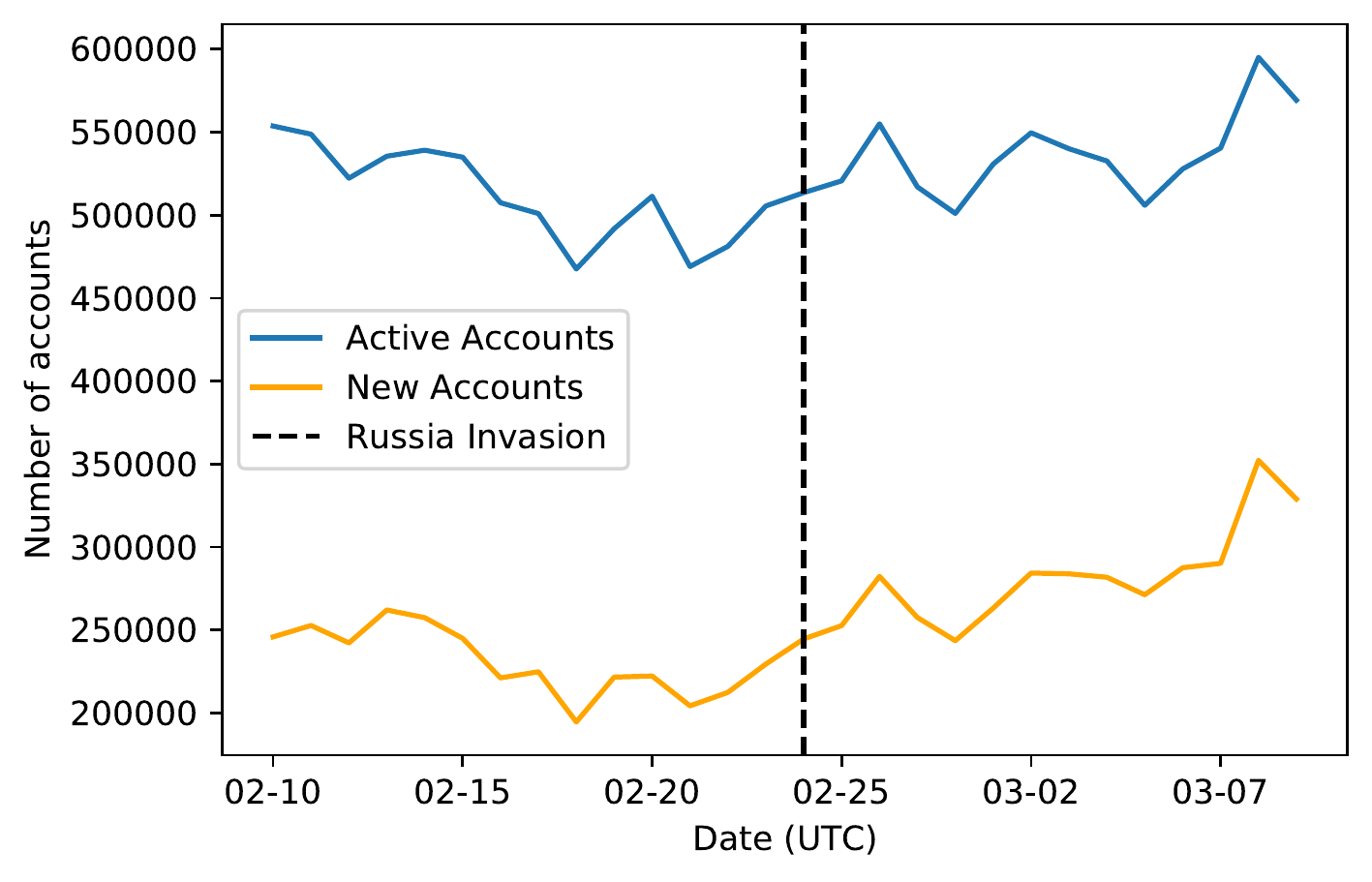}\label{subfig:accmain}}
    \subfigure[Flashbots accounts]{\includegraphics[page=1,width=.52\textwidth]{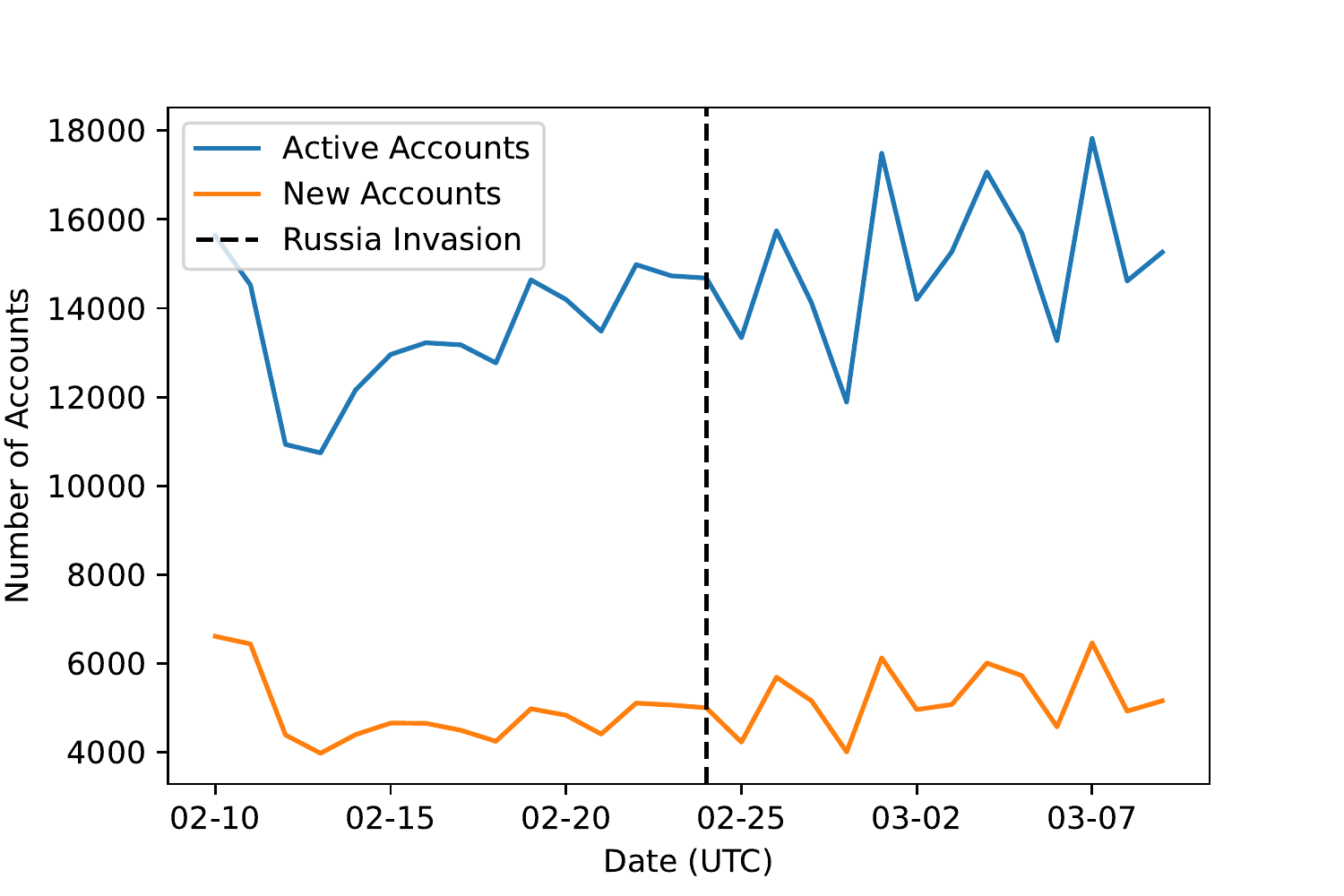}\label{subfig:accflashbots}}
    \caption{Active accounts and new accounts.}
    \label{fig:accfig}
\end{figure}

In this topic, we present the results for the analysis of account activity. As discussed in Section \ref{sec:modelagem}, we model the network as a oriented graph in which each account corresponds to a node. From this, we built time charts to relate the activity of these nodes over the days of the analyzed period, contrasting the volume of active accounts with the occurrence of new accounts. All results are shown in Figure \ref{fig:accfig}.

For the set $C_1$, Figure \ref{subfig:accmain}, we observed that the activity of the accounts remained stable, even though we observed a decrease in active addresses between the 10th and 20th of February, keeping the volume of active nodes between 450 thousand and 550 thousand. After that, the curve starts to have a growth trend that started close to the invasion that extends until the beginning of March. Similarly, the volume of new accounts follows the same pattern, varying between 200,000 and 300,000 new accounts in mid-February and later being characterized by growth until the end of the analysis period, with a more accentuated variation close to the 10th of February. March when the volume of new nodes approaches 350 thousand.

Regarding the $C_2$ set, Figure \ref{subfig:accflashbots}, it is noted that despite the occurrence of a fall of about 5 thousand accounts in relation to the number of active nodes between February 10th and 15th, the graph maintains the trend of growth over time, with more intense variations from the day of the invasion. In this context, it is noted that, as for the $C_1$ set, the new accounts curve follows a similar pattern to the active addresses curve, with the former showing a significantly smoother growth than the latter, maintaining the variation at approximately a thousand addresses.

\subsection{Main Accounts}~\label{subsec:topaccs}
\begin{figure}[!ht]
      	\includegraphics[page=1,width=\textwidth]{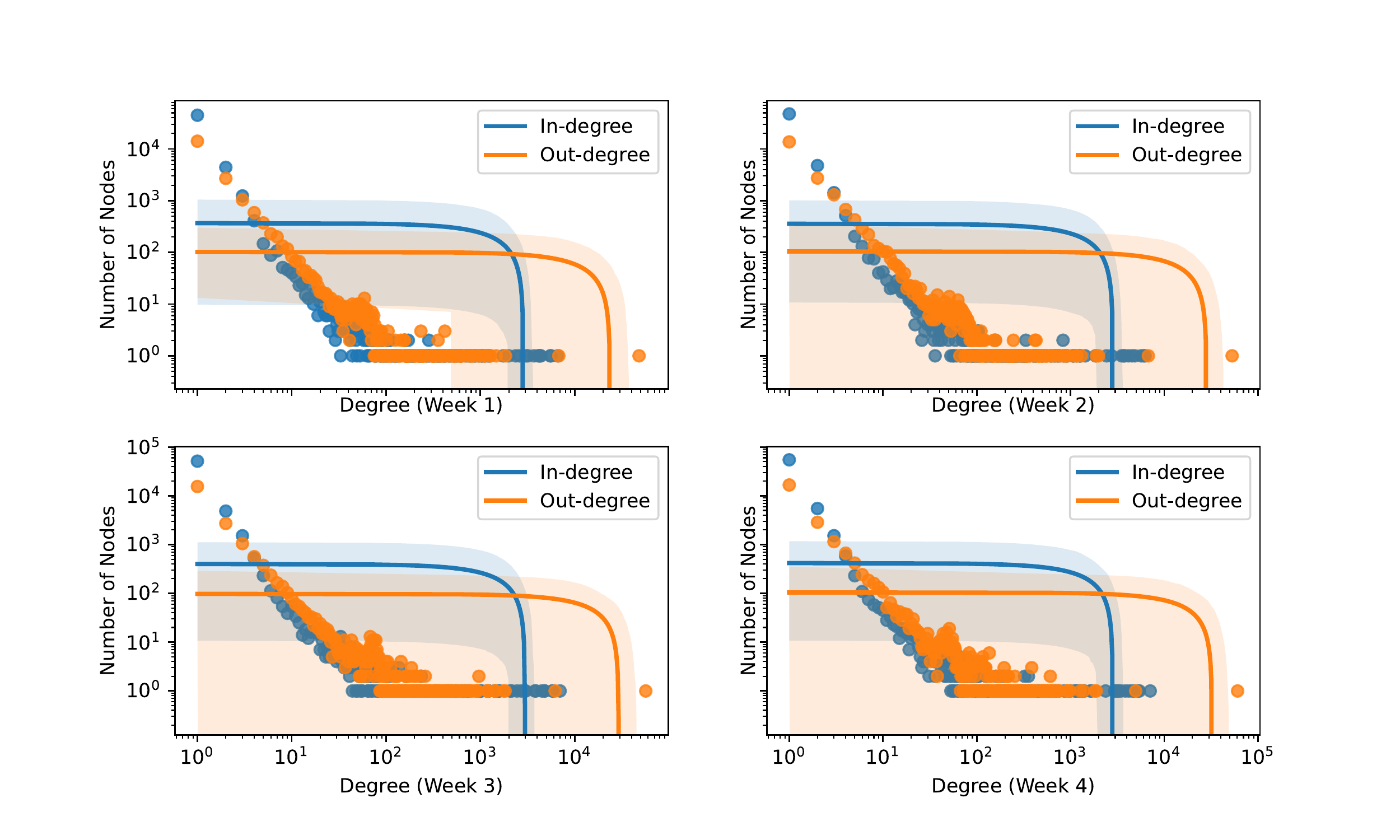}
        \caption{Relation of degree between weeks for Flashbots}
        \label{fig:degreeweek_flashbots}
\end{figure}

\begin{figure}[!ht]
      	\includegraphics[page=1,width=\textwidth]{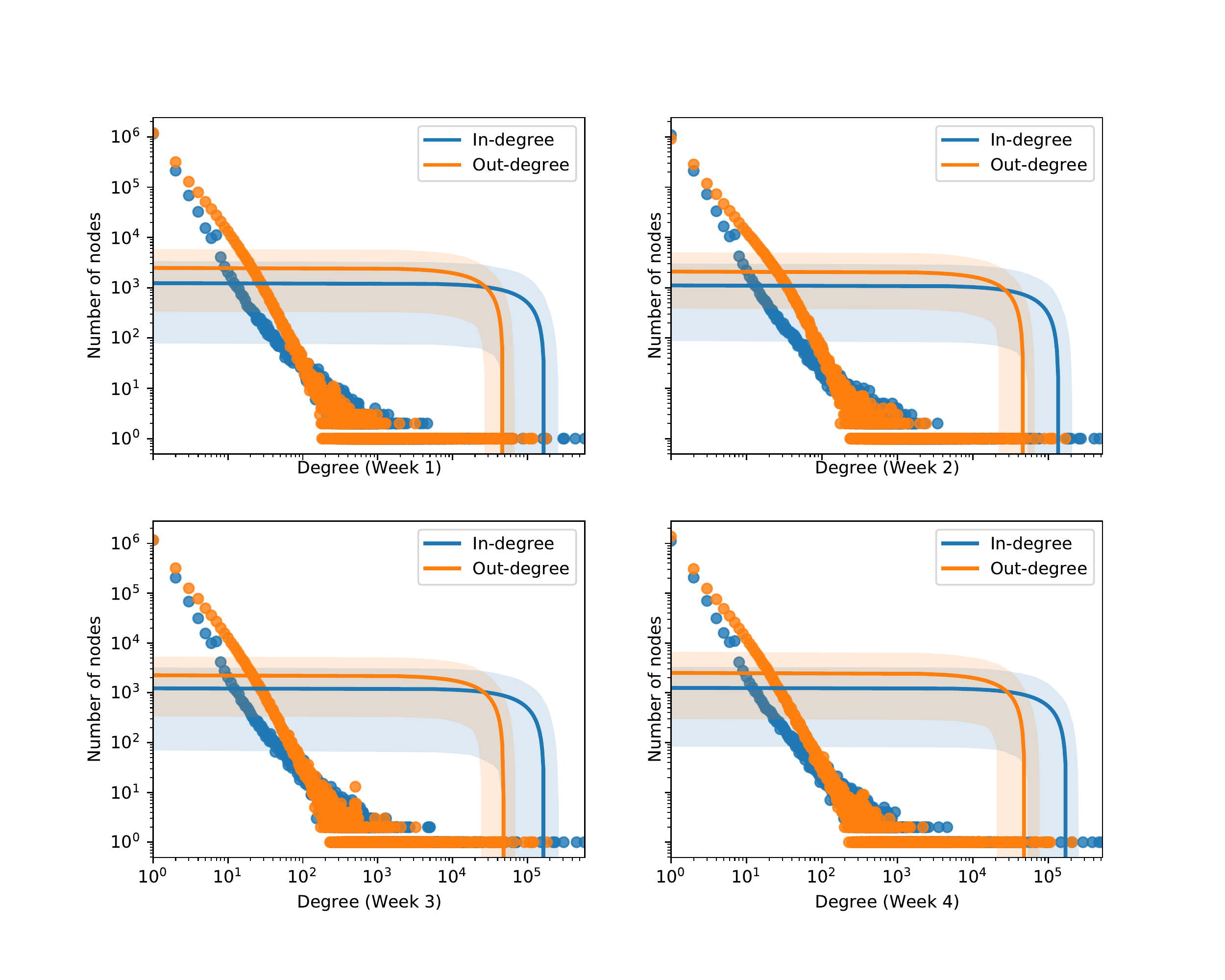}
        \caption{Relation of degree between weeks for mainnet Ethereum}
        \label{fig:degreeweek_main}
\end{figure}

As discussed in Section \ref{subsec:mainacc}, we model the network through directed graphs. Figures \ref{fig:degreeweek_flashbots} and \ref{fig:degreeweek_main} show the distribution of values of different inflows and outflows between accounts during each week. The higher volume of transactions in the main network database decreases uncertainty within this distribution.


Note that the behavior of the blue and orange curves in the graphs in Figure \ref{fig:degreeweek_flashbots} show that some accounts, in a small quantity, have higher outdegree, while the indegree curve does not reach such high values. Thus, the profile of this database in this period indicates the presence of accounts distributing many transactions to different accounts and many accounts receiving few transactions. This behavior does not change significantly over the weeks.

Observing Figure \ref{fig:degreeweek_main}, this pattern between the curves is reversed. In this case, there appears to be a greater number of accounts with a high indegree, which reflects slightly different behavior, with more accounts receiving transactions from many different users. It is important to emphasize that in this case, the curves are closer to each other, which indicates a more balanced behavior in this sense. Likewise, the passing of weeks did not significantly change the behavior of the network.

In addition to a general observation about the grades of the accounts, we create a ranking in order to find accounts considered relevant for the period in question. The relevance of an account, in this work, is defined by how much the degree of an account increases from one week to another, that is, with how many accounts it started to send and receive transactions.

That said, we performed a four-week comparative analysis, comparing the accounts that increased their grade the most between one week and the following week. In order to compress the results, we will only show the 10 most relevant accounts in the period. Additionally, hash addresses for each account will be shortened to the last seven characters of the address. For example, the account \textit{0xa090e606e30bd747d4e6245a1517ebe430f0057e} will only be shown as \textit{0f0057e}. To facilitate the identification of accounts, we label each account according to publicly available tags on Etherscan.

Within this collection period, the first two weeks describe the period before the Russian invasion. Possible effects of pre-conflict tension may be reflected within this period. Weeks 3 and 4 describe the period after the invasion, which may contain the effects and consequences of the new situation.

The first analyzed interval was the interval of the first two weeks. It is observed that, as Table \ref{tab:mainaccs1} demonstrates, a large number of accounts related to token trading stand out. X2Y2\footnote{\url{https://x2y2.io/}}, for example, is a \textit{marketplace} for NFT exchanges and trades. This also occurs in the context of Flashbots. As shown in Table \ref{tab:flashbotsaccs1}, the account with the highest degree growth was OpenSea\footnote{\url{https://opensea.io/}}, which also deals with a market focused on trading tokens. In addition, the presence of a wallet address without a public tag on Etherscan is noted, which obtained a degree increase of 1200 units, as well as the occurrence of other assets such as USDT Stablecoin with Tether\footnote{\url{https:// tether.to/}} and DEX aggregator with 1inch\footnote{\url{https://app.1inch.io/}}. It is also worth noting the appearance of MEV extraction bots as relevant addresses, given the purpose of Flashbots as a safe and transparent environment for the extraction of these recipes.

\begin{table}[!ht]
\centering
\caption{Comparasion of accounts between the first and second weeks in the mainnet of Ethereum.}
\label{tab:mainaccs1}
\begin{tabular}{ccc}
\toprule
Address & Tag             & Degree Growth \\
\midrule
1abebc9           & X2Y2: X2Y2 Token                   & 44774                                            \\
49a41fa           & X2Y2: X2Y2 Drop                    & 27772                                            \\
a67eed3           & X2Y2: Exchange                     & 24353                                            \\
b23c98a           & Cat Blox: CATBLOXGEN Token         & 17192                                            \\
c756cc2           & Wrapped Ether                      & 15267                                            \\
76a1b85           & X2Y2: Fee Sharing System           & 14792                                            \\
147ea85           & ENS: Base Registrar Implementation & 12849                                            \\
903a5d0           & The Sandbox: SAND Token            & 12720                                            \\
5fd74c0           & \textit{Contract: No public tag} & 12403                                            \\
def3c7a           & \textit{Contract: No public tag} & 11541                                            \\
\bottomrule
\end{tabular}
\end{table}
\begin{table}[!ht]
\centering
\caption{Comparasion of accounts between the first and second weeks in the Flashbots subset.}
\label{tab:flashbotsaccs1}
\begin{tabular}{ccc}
\toprule
Address & Tag             & Degree Growth \\
\midrule
bB538E5           & OpenSea: Wyvern Exchange v2       & 3787                                            \\
5843dE9          & \textit{Usuário: Sem tag pública}  & 1200                                            \\
991e2D4           & MEV Bot: 0x000...2D4              & 833                                            \\
7455580           & MEV Bot: 0x8aF...580              & 768                                            \\
57Fb793           & Wintermute 1                       & 597                                            \\
643097d           & 1inch v4: Router                   & 335                                            \\
140449c           & \textit{Contract: No public tag} & 325                                            \\
5206C3E           & \textit{Contract: No public tag} & 287                                            \\
3f7c613           & \textit{Contract: No public tag} & 282                                            \\
D831ec7           & Tether: USDT Stablecoin            & 259                                            \\
\bottomrule
\end{tabular}
\end{table}

Then, a comparison was made between weeks 2 and 3 of the collection period. This analysis allows us to observe transformations the Ethereum network had during the period of the Russian invasion, comparing the week immediately before and the first week after the fact. It can be noted, comparing Table \ref{tab:mainaccs1} with Table \ref{tab:mainaccs2}, that the previously high quantity of tokens gave rise to a greater variety of account profiles. Coinbase exchange starts to stand out a little more. There is also a greater presence of contracts related to other assets, such as USDT Stablecoin or USD Coin. Finally, the account \textit{7455580} also displays a bot aimed at MEV extraction activities. 

Regarding Flashbots, comparing the Tables \ref{tab:flashbotsaccs1} and \ref{tab:flashbotsaccs2}, it is possible to highlight the increase in the number of MEV bots among the most relevant nodes, as well as the appearance of addresses without public tags on the network. Furthermore, it is worth noting that Ethermine\footnote{\url{https://ethermine.org/}} appears in the first position in Table \ref{tab:flashbotsaccs2} as the node with the highest degree growth, with this growth being almost 55 thousand units higher than the account that occupies the second position in the table.

\begin{table}[!ht]
\centering
\caption{Comparasion of accounts between the second and third weeks in the mainnet of Ethereum.}
\label{tab:mainaccs2}
\begin{tabular}{ccc}
\toprule
Address & Tag             & Degree Growth \\
\midrule
0f0057e           & Coinbase: Miscellaneous           & 106853                                           \\
bb538e5           & OpenSea: Wyvern Exchange v2       & 101837                                           \\
d831ec7           & Tether: USDT Stablecoin           & 48732                                            \\
606eb48           & Centre: USD Coin                  & 35541                                            \\
8d01279           & Celsius Network: Wallet 5         & 23776                                            \\
7975ef8           & OogaVerse: MekaApes Game Contract & 16276                                            \\
b898ec8           & Ethermine                         & 15246                                            \\
7455580           & MEV Bot: 0x8aF...580              & 13650                                            \\
dc49699           & Coinbase 4                        & 13600                                            \\
bf77006           & \textit{Contract: No public tag} & 12533                                            \\
\bottomrule
\end{tabular}
\end{table}
\begin{table}[!ht]
\centering
\caption{Comparasion of accounts between the second and third weeks in the Flashbots subset.}
\label{tab:flashbotsaccs2}
\begin{tabular}{ccc}
\toprule
Address & Tag             & Degree Growth \\
\midrule
B898ec8           & Ethermine                         & 56624                                           \\
991e2D4           & MEV Bot: 0x000...2D4              & 1881                                           \\
0a3f8e7           & MEV Bot: 0x4Cb...8e7              & 1568                                            \\
C378B9F           & SushiSwap: Router                 & 1324                                            \\
5645718           & \textit{User: No public tag} & 1195                                            \\
e416B40           & MEV Bot: 0x000...B40              & 1191                                            \\
0fDd6CF           & MEV Bot: 0xa57...6CF              & 1117                                            \\
7455580           & MEV Bot: 0x8aF...580              & 724                                            \\
01Db497           & \textit{User: No public tag}  & 703                                            \\
08850aF           & \textit{User: No public tag}  & 666                                            \\
\bottomrule
\end{tabular}
\end{table}

The two last weeks form the last analysis window, comparing weeks after the invasion. Looking at Table \ref{tab:mainaccs3}, some characteristics remain compared with Table \ref{tab:mainaccs2}, such as the presence of MEV extraction bots and the still high relevance of accounts such as Ethermine and Coinbase: Miscellaneous. Token accounts reappear with relevance, as in the first two weeks, which can be seen, for example, with the growth of the Uniswap protocol, which can be used to exchange and acquire tokens\footnote{\url{ https://uniswap.org/faq}}. Another new observation in this window is the presence of accounts with greater relevance that do not have public identification tags on Etherscan. The same behavior is observed with Flashbots which, as shown in Table \ref{tab:flashbotsaccs3}, keep MEV bots among the nodes with the highest degree growth, in addition to the greater relevance of nodes focused on token trading, such as the previously mentioned Uniswap.

\begin{table}[!ht]
\centering
\caption{Comparasion of accounts between the third and fourth weeks in the mainnet of Ethereum.}
\label{tab:mainaccs3}
\begin{tabular}{ccc}
\toprule
Address & Tag             & Degree Growth \\
\midrule
33ab239           & Livepeer: LPT Token      & 95535                                            \\
0f0057e           & Coinbase: Miscellaneous  & 76216                                            \\
97b3fca           & \textit{Contract: No public tag} & 56225                                 \\
665fc45           & Uniswap V3: Router 2     & 52794                                            \\
b898ec8           & Ethermine                & 21609                                            \\
9f2488d           & Uniswap V2: Router 2     & 19930                                            \\
500fb08           & \textit{Contract: No public tag} & 17267                                 \\
d831ec7           & Tether: USDT Stablecoin  & 16330                                            \\
0a3f8e7           & MEV Bot: 0x4Cb...8e7     & 12988                                            \\
6155ba9           & KPOP CTzen: KCPT1 Token  & 11093                                            \\
\bottomrule
\end{tabular}
\end{table}
\begin{table}[!ht]
\centering
\caption{Comparasion of accounts between the third and fourth weeks in the Flashbots subset.}
\label{tab:flashbotsaccs3}
\begin{tabular}{ccc}
\toprule
Address & Tag             & Degree Growth \\
\midrule
B898ec8           & Ethermine                & 3950                                            \\
0a3f8e7           & MEV Bot: 0x4Cb...8e7     & 2986                                            \\
40f594e           & MEV Bot: 0x000...94e     & 1398                                            \\
9F2488D           & Uniswap V2: Router 2     & 1306                                            \\
3923BcA           & \textit{User: No public tag} & 922                                    \\
9e5D3Fa           & \textit{Contract: No public tag} & 636                                            \\
80E83b6           & Gelato Network: Gelato   & 444                                             \\
665Fc45           & Uniswap V3: Router 2    & 421                                             \\
643097d           & 1inch v4: Router         & 409                                             \\
A7fAc82           & \textit{User: No public tag} & 333                                  \\
\bottomrule
\end{tabular}
\end{table}

\section{Conclusion}~\label{conclusao}
In this work, we characterize the Ethereum network in a general and specific way, building an evolutionary model of the network and performing quantitative and qualitative analyzes on its accounts. As a result, we sought to quantify the active nodes on the network and, finally, isolated and analyzed the most relevant accounts during that period.

Our investigations indicate an initial path for monitoring the possible effects of important external events on a large cryptocurrency platform. It was possible to observe short-term changes concerning the behavior of accounts and transactions on the network, in addition to the contrast between the behavior of transactions on the general network and transactions carried out with Flashbots Auction.

It is important to emphasize that this work seeks only to make a characterization concentrated in a period and that these analyses verify a short-term behavior. Therefore, it is not possible to make cause-and-effect inferences regarding the invasion just by looking at these specific details.

For future works, we intend to use wider temporal windows and analyze in more detail the most influential nodes to enable more assertive causal inferences. Another focal point is to expand the investigation of the impact of external events on other past events that have economic relevance.

\section*{Acknowledgments}

The authors acknowledge the financial support of CNPq, FAPEMIG, Capes and UFJF.

\bibliographystyle{comnet}
\bibliography{paper}

\end{document}